# Substrate induced nanoscale resistance variation in epitaxial graphene

A. Sinterhauf,[1] G. A. Traeger,[1] D. Momeni Pakdehi,[2] P. Schädlich,[3] P. Willke,[4,5] F. Speck,[3] T. Seyller,[3] C. Tegenkamp,[3] K. Pierz,[2] H. W. Schumacher,[2] M. Wenderoth[1]

[1] *IV. Physikalisches Institut, Georg-August-Universität Göttingen, Friedrich-Hund-Platz 1, 37077 Göttingen, Germany*

[2] *Physikalisch-Technische Bundesanstalt, Bundesallee 100, 38116 Braunschweig, Germany*

[3] *Institut für Physik, Technische Universität Chemnitz, Reichenhainer Straße 70, 09126 Chemnitz, Germany*

[4] *Center for Quantum Nanoscience, Institute for Basic Science (IBS), Seoul 03760, Republic of Korea*

[5] *Department of Physics, Ewha Womans University, Seoul 03760, Republic of Korea*

E-Mail: anna.sinterhauf@uni-goettingen.de
martin.wenderoth@uni-goettingen.de

## Abstract

Graphene, the first true two-dimensional material still reveals the most remarkable transport properties among the growing class of two-dimensional materials. Although many studies have investigated fundamental scattering processes, the surprisingly large variation in the experimentally determined resistances associated with a localized defect is still an open issue. Here, we quantitatively investigate the local transport properties of graphene prepared by polymer assisted sublimation growth (PASG) using scanning tunneling potentiometry. PASG graphene is characterized by a spatially homogeneous current density, which allows to analyze variations in the local electrochemical potential with high precision. We utilize this possibility by examining the local sheet resistance finding a significant variation of up to 270% at low temperatures. We identify a correlation of the sheet resistance with the stacking sequence of the 6H-SiC substrate as well as with the distance between the graphene sheet and the substrate. Our results experimentally quantify the strong impact of the graphene-substrate interaction on the local transport properties of graphene.

**Introduction**

Charge transport in graphene has been subject of intense theoretical and experimental investigation since its first electronic characterization [1]. The high quality and its 2D nature make epitaxial graphene the perfect system to study fundamental transport properties on the nanometer scale. Consequently, in a series of experiments - based on scanning tunneling potentiometry (STP) [2] or 4-point-probe microscopy [3] - several groups have focused on the impact of local scattering centers like single substrate steps [4, 5] or the transition from monolayer to bilayer graphene on transport [6, 7]. From these results, it is qualitatively well understood that the transport properties of epitaxial graphene samples are not homogeneous on the nanometer scale. Substrate steps or monolayer-bilayer junctions, which are typically present in such samples, act as local scattering centers. In addition, for epitaxial graphene on SiC(0001) it is well known that interaction with the substrate drastically affects the transport properties of the graphene sheet. In order to reduce this inherent proximity effect, i.e. to improve the transport properties of the graphene sheet, different strategies were pursued such as the refinement of the growth process [8, 9, 10], the use of suitable dielectric substrates like boron nitride [11], the decoupling of the substrate by intercalation [12], or the preparation of suspended graphene [13]. Moreover, the proximity effect can also be deliberately exploited to specifically tune the properties of a graphene sheet [14, 15, 16, 17, 18]. For example, the almost negligible spin-orbit coupling in graphene can be significantly increased by bringing the graphene layer into contact with transition metal dichalcogenides [14, 15]. The deposition of ferromagnetic materials on graphene results in proximity-induced ferromagnetic correlations [17] and proximity superconductivity can be observed in graphene in the vicinity of superconducting materials [19].

In the context of charge transport in epitaxial graphene, local defects like substrate steps as well as local variations of the coupling between the graphene layer and the substrate result in a locally varying potential landscape as well as a spatially inhomogeneous current density. Analyzing the published results for the resistances assigned to specific defects present in epitaxial graphene, one finds a large spread [4, 5, 6, 7, 20, 21, 22, 23, 24, 25]. The strong variation in the experimental values of sheet or defect resistances determined by local probe measurements is very likely due to the lack of information about the actual local current density. Replacing the probe by a single-electron transistor allows simultaneous measurement of local voltage drop and current distribution in 2D materials [26] with a lateral resolution in the range of 350 nm [26]. In comparison, STP has an angstrom resolution [27] and measures the local electrochemical potential with high accuracy, but local variations in the current density are experimentally not accessible and are indistinguishable from spatial variations of the sheet resistance. For conventionally grown graphene on SiC(0001), typically monolayers as well as bilayers are present. Monolayer-bilayer transitions represent a strong scattering center and cause a significant variation of the local current density. Having no better approach, so far local transport properties have been determined using an averaged (sometimes even macroscopic) current density for the whole sheet. In this study, we show that the high quality of polymer-assisted epitaxial monolayer graphene samples opens a promising way to quantify local transport properties with high precision. Applying the PASG method, it is possible to grow large-scale monolayer graphene sheets [9, 28] and especially to completely suppress the formation of bilayer areas. In 4-terminal electronic transport measurements, a remaining resistance anisotropy of only 3% was reported that was traced back to the residual resistance caused by substrate steps [28]. Here, we investigate on the nanometer scale spatial variations of the interaction between the substrate of the PASG samples and a perfect monolayer graphene sheet. We quantify the sheet resistance $\varrho_{\text{sheet}}$ with high precision, and reveal a significant variation of $\varrho_{\text{sheet}}$ connected with the distance between the graphene layer and the substrate as well as with the stacking of the 6H-SiC substrate. Furthermore, analyzing the temperature-dependence of the sheet resistance we report a large inhomogeneity in the sheet resistance at low temperatures and discuss different scattering mechanisms.

## Experimental results

### Homogeneity of the current density

The local electric field as well as the local current density are needed to determine the local sheet resistance. While STP measures the local voltage drop, the local current density is a priori unknown. In STP studies, it is replaced by an averaged value, e.g. given by the total current and the geometry of the sample. While this approximation has severe limits for locally inhomogeneous samples, the excellent lateral homogeneity of the PASG graphene parallel to the steps, the absence of bilayer graphene and the low impact of steps on the overall resistance [28], drastically reduce lateral current density variations [29]. In our STP setup, the current flow was deliberately driven parallel to the miscut of the SiC sample, resulting in an overall voltage drop perpendicular to the substrate steps ([1$\bar{1}$00] direction). The geometry chosen in the experiment and the assumption that graphene terraces have a constant sheet resistance parallel to the steps result in a constant average current density on all terraces. To estimate the remaining variation in $j_{\text{local}}(x, y)$ we have modelled the local current density for a the given surface geometry taken from constant current topographies (CCT) with a resistor network [6, 22, 30]. With a given surface morphology (e.g. Fig.1a), the resulting current density exhibits a maximum variation of up to 7% (Fig. 1b). By carefully selecting regions away from complex step configurations, such as regions where two steps converge, the current density can be considered as highly homogeneous. It is $j_{\text{local}}(x,y) \approx j_{\text{local}} = (0.89 \pm 0.01)\text{Am}^{-1}$ for an applied voltage along the graphene layer perpendicular to the substrate steps of 1V at $T = 300$ K. The absolute values of the current density in the selected regions are given in Supplementary Table 1. Comparing the lateral variation of the current density in PASG graphene samples with conventionally grown epitaxial graphene, it becomes obvious that local variations in $\varrho_{\text{sheet}}$ from mono- and bilayer graphene and monolayer-bilayer junctions in conventionally grown epitaxial graphene induce a strong variation of $j_{\text{local}}(x,y)$ (Supplementary Fig. 1).

Fig. 1

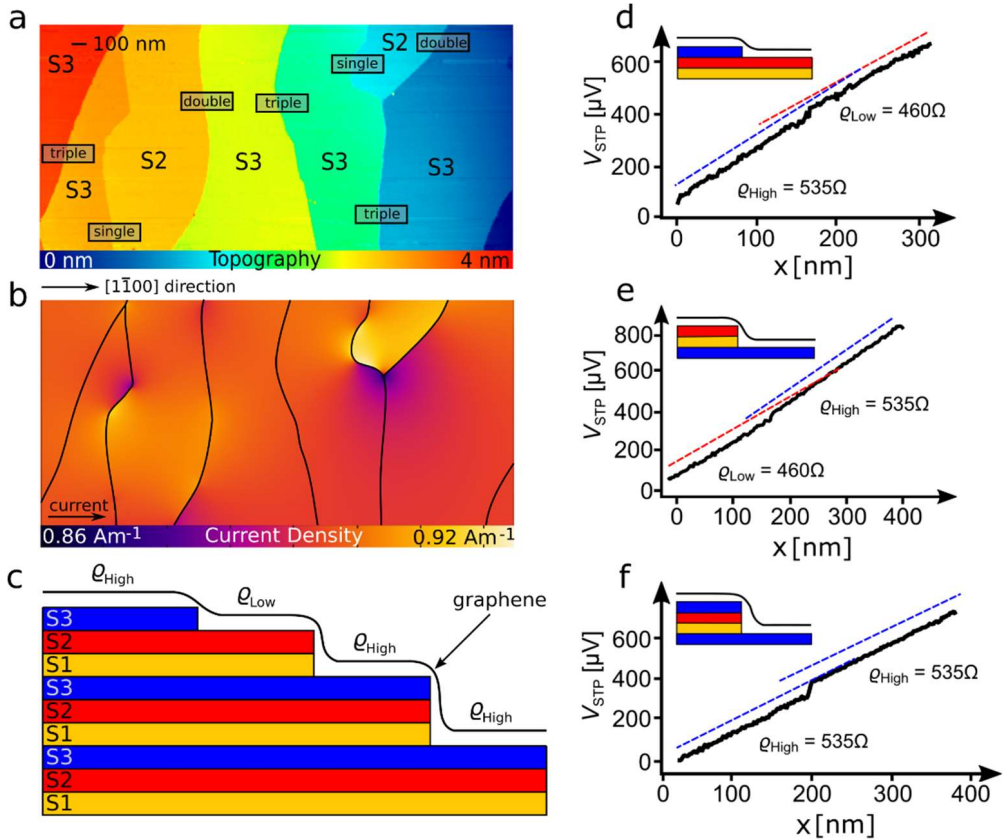

Current density and evaluation of the local sheet resistance at room temperature. **a** large-scale constant current topography (2μm x 1μm, $V_{\text{Bias}} = 0.5V, I_T = 0.03nA$). S1,

S2 and S3 indicate the fundamental bilayers (and thus the stacking) of the SiC substrate, details are given in the discussion. STP measurements were performed in the marked areas (black boxes). The height of the steps is denoted in the marked areas. Using the macroscopic ohmic resistance, the sample geometry shown in a and step resistances of $6\,\Omega\mu m$, $12\,\Omega\mu m$, $18\,\Omega\mu m$ for single, double and triple steps, respectively, **b** the $[1\bar{1}00]$ component of the local current density $j_{\text{local}}(x,y)$ is calculated with a finite element simulation. **c** schematic side view of the crystal structure of 6H-SiC. **d** measured voltage drop along the graphene layer induced by the cross voltage $V_{\text{cross}}$ when crossing a single step. Dashed lines represent the slope of the voltage drop (shifted for clarity). **e** corresponding measurement for a double step and **f** for a triple step.

### Local variation of $\varrho_{\text{sheet}}$ at temperature $T = 300K$

Large scale constant current topographies (Fig.1a) reveal a surface with single, double as well as triple substrate steps and no bilayer regions as expected for epitaxial graphene grown by PASG [9, 10, 28]. STP measurements investigating $\varrho_{\text{sheet}}$ are performed across all step configurations in Fig. 1a, the corresponding voltage drops $V_{\text{STP}}$ are shown in Fig. 1d-f. Interestingly, to the left and to the right of single substrate steps we find a different gradient of $V_{\text{STP}}$ (Fig. 1d), indicated by the dashed blue and red lines representing the slope to the left and to the right of the step, respectively. Since the current density is constant, this directly proves that the top and bottom terrace have different sheet resistances $\varrho_{\text{sheet}}$. This finding also holds for terraces connected by a double substrate step (Fig. 1e), whereas the identical $\varrho_{\text{sheet}}$ is measured when crossing a triple substrate step (Fig. 1f). For all step configurations an additional voltage drop at the topographic position of the step is observed, which is commonly explained by a potential barrier induced by the step due to detachment of the graphene sheet from the substrate [4, 5, 31].

In order to further investigate spatial variations of the sheet resistance, we have measured large sequences of steps. The topographic analysis has shown that instead of a random distribution of step heights, a well-defined sequence of the step heights shows up: along the $[1\bar{1}00]$ direction, either a triple substrate step is present or a single substrate step and a double substrate step are observed. These characteristic step patterns for PASG graphene on 6H-SiC have recently been reported in an Atomic Force Microscopy (AFM) study and have been attributed to the growth process [28]. The detailed STP analysis of large sequences of substrate steps allows deducing two important implications: firstly, the evaluation shows that at 300K the sheet resistance across a given terrace is constant (Supplementary Fig. 2). Secondly, from STP measurements on more than 40 terraces, we extract two clearly distinct sheet resistances, which we refer to as $\varrho_{\text{High}}$ and $\varrho_{\text{Low}}$. The mean $\varrho_{\text{High}}$ is 535Ω and for the mean $\varrho_{\text{Low}}$ we find 460Ω. The mean $\varrho_{\text{High}}$ and $\varrho_{\text{Low}}$ deviate by $(14 \pm 1)\%$ from each other at room temperature. Moreover, $\varrho_{\text{High}}$ as well as $\varrho_{\text{Low}}$ show a variation from terrace to terrace of $\pm 20\Omega$.

**Temperature-dependence of $\varrho_{\text{sheet}}$**

In order to disentangle possible scattering processes and to understand the difference between $\varrho_{\text{High}}$ and $\varrho_{\text{Low}}$, we performed further temperature-dependent STP measurements at 77K and 8K (Fig. 2a). We find an overall decrease in the sheet resistance with decreasing temperature, which is supported by macroscopic transport measurements in 4-point van der Pauw geometry (Supplementary Fig. 3) and in agreement with published results [32, 33]. The relative reduction in $\varrho_{\text{High}}$ with decreasing temperature is slightly smaller, i.e. it reduces by 32%, from 535Ω to approximately 365Ω at 8K, compared to the temperature-dependence of $\varrho_{\text{Low}}$, which declines from 460Ω to an average value of 250Ω at 8K, i.e. it reduces by 45%. Besides the overall reduction of the sheet resistance, a surprising large increase in the spread in the data is observed with decreasing temperature. At 8K, a maximum variation in $\varrho_{\text{sheet}}$ of approximately 270% between the lowest value for $\varrho_{\text{Low}}$ and the highest value for $\varrho_{\text{High}}$ is observed.

Fig. 2

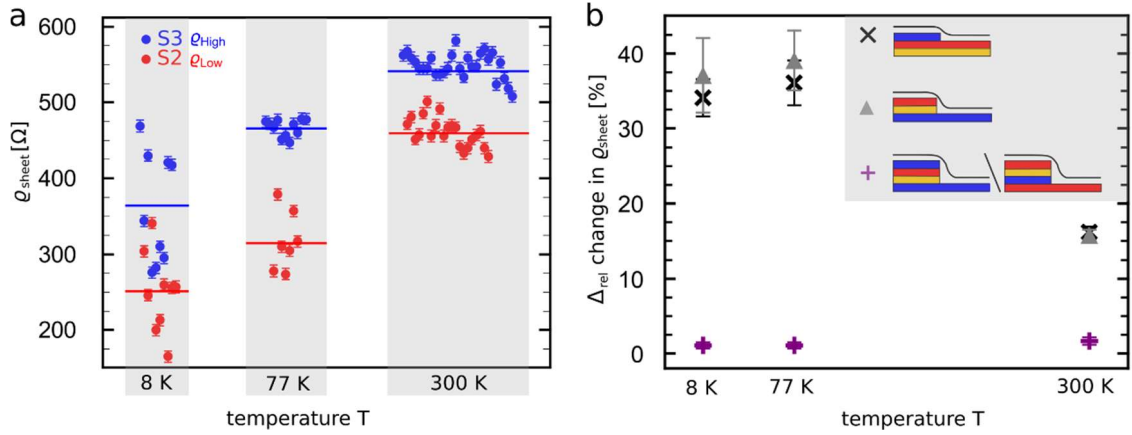

Temperature-dependence of the sheet resistance. **a** sheet resistance at 8K, 77K and 300K acquired on more than 80 terraces, solid horizontal lines indicate the mean value for a given terrace and temperature. **b** change in sheet resistance for adjacent terraces for all three cases S3→S2, S2→S3 and S3→S3 / S2→S2 as a function of temperature.

On adjacent terraces a maximum variation of 178% (Supplementary Fig. 4) is measured. In the following we will use $\Delta_{\text{rel}} = (\varrho_{\text{sheet1}} - \varrho_{\text{sheet2}})/(\varrho_{\text{sheet1}} + \varrho_{\text{sheet2}})/2$ to quantify the relative change in the sheet resistance for adjacent terraces (Fig. 2b). Regardless of the temperature, when crossing a triple substrate step, the variation in $\varrho_{\text{sheet}}$ is small, i.e. $\Delta_{\text{rel}} < 3\%$. In contrast to this, the relative variation in $\varrho_{\text{sheet}}$ to the left and to the right of single as well as double substrate steps drastically increases when going from 300K to 77K. In particular, for terraces connected by single or double substrate steps a relative change of more than 30% is measured. In both cases $\Delta_{\text{rel}}$ slightly decreases from 77K to 8K.

### Analysis of the surface morphology of individual steps and terraces

To further investigate the local variation of the transport properties, structural and electronic properties of PASG graphene have been analyzed on different length scales on the same samples. On a mesoscopic scale the surface is characterized by single, double and triple steps, resulting from the miscut of the SiC substrate. Surprisingly, we rarely found the expected height of the substrate steps, i.e. multiple of $0.25$ nm [34]. Instead, we observed deviations of the step height with smaller as well as larger values for both single and double steps. As an example, Fig. 3a displays a line profile across a step sequence consisting of a single substrate step and a double substrate step.

Fig. 3

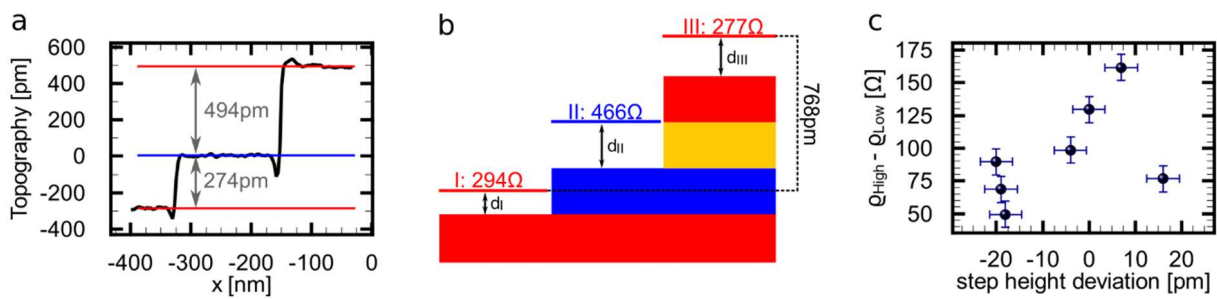

Analysis of the step height of single and double steps. **a** Line profile through a constant current topography showing adjacent terraces S2, S3, S2, connected by a single substrate step followed by a double substrate step recorded at $77$K. The line profile reveals a deviation from the step heights of the SiC substrate steps. **b** schematic representation of the correlation between step height and sheet resistance illustrating a locally varying distance d between the graphene layer and the substrate. **c** difference in the sheet resistance for adjacent terraces for single and double steps measured at $8$K as a function of the deviation of the step height. Details are given in the discussion.

For this specific step configuration, the analysis reveals a step height $> 0.25$nm for the single substrate step and a step height $< 0.5$nm for the double substrate step. It is worthwhile to mention, that also the combined step height does not fit to the expected value of three times $0.25$ nm. Assuming that different step heights correspond to different distances between the graphene monolayer and the substrate, such step sequences allow to study the relation between distance and sheet resistance. The corresponding STP measurement reveals that the terrace III (see Fig. 3b) has a higher conductivity compared to terrace I, indicating that a larger distance results in a smaller resistance. Details on the dependence of the sheet resistance on the step height are summarized in Supplementary Fig. 5, the general trend supports the assumption that larger distances result in higher conductivities. Moreover, comparing terraces connected by steps with almost identical step height (e.g. Supplementary Fig. 5 black: 507pm and pink: 500pm), we find a large spread of the sheet resistances: 304Ω vs. 465Ω and 165Ω vs. 294Ω for the black and pink configuration, respectively. The topographic line profiles as well as the corresponding dI/dV-spectra for the two data sets are given in Supplementary Fig. 6a-c. Height deviations are found at 8K (Supplementary Fig. 5), 77K (Fig. 3a) and 300K (Supplementary Fig. 6d, e) and the topographic nature of the observed height deviation in CCT is supported by AFM topographies (Supplementary Fig. 7). Details of the height analysis are summarized in Supplementary Fig. 8; height calibration (Supplementary Fig. 9) was done at an adjacent triple substrate step exhibiting the same sheet resistance on both terraces.

In order to take the atomic scale structure of the sample into account, we acquired higher resolved CCTs on terraces connected by single and double substrate steps as shown in Fig. 4a and Supplementary Fig. 10b, respectively. On all terraces the $6 \times 6$-quasi corrugation is clearly visible. This modulation is induced by a lattice mismatch of the graphene sheet and the substrate and originates from actual height corrugation as well as from electronic contrast [35, 36, 37, 38]. However, this modulation is structurally not perfect (compare Fig. 4a). In order to analyze deviations from a perfect ordering, we disentangle the constant current topographies using Fourier analysis (Supplementary Fig. 11). Applying this type of evaluation for each terrace separately, we disentangle three different contributions to the topographic contrast. Firstly, the $6 \times 6$-quasi corrugation itself, secondly short-range noise and thirdly, long-range spatial modulations, which can be understood as perturbations of the $6 \times 6$-quasi corrugation.

The latter contributions are shown in Fig. 4b, c for the terraces to the left and to the right of the single substrate step in Fig. 4a, respectively. We determine the dominant wavelength of these modulations as shown in Supplementary Fig. 12. A clear difference between the two terraces can be identified. The terrace to the left of the single substrate step (Fig. 4b) shows a spatial modulation with a shorter wavelength of $4.2$nm compared to the wavelength of the spatial modulations on the terrace to the right of the single substrate step (Fig. 4c) with $8.1$nm. The corresponding, i.e. reversed finding, holds for terraces connected by a double substrate step (Supplementary Fig. 10b) for which comparable wavelengths of the spatial modulations are extracted. Besides differences in the dominant wavelength, the spatial modulations also exhibit different amplitudes.

In summary, the analysis of the surface morphology allows two important conclusions. Firstly, the deviation of the step heights indicates a locally varying distance between the graphene layer and the substrate. Secondly, the $6 \times 6$-quasi corrugation does not show a perfect ordering.

Fig. 4

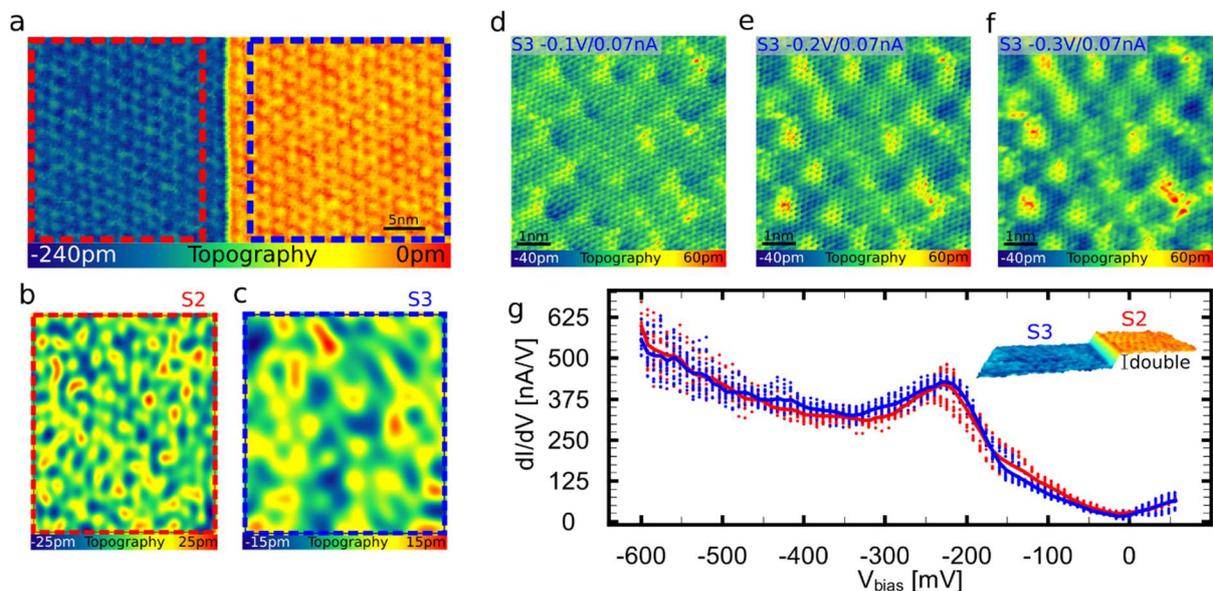

Analysis of the local defect structure on terraces S2 and S3. **a** 50 nm x 25 nm constant current topography of terraces connected by a single substrate step. On both terraces the $6 \times 6$ modulation is well resolved. The topographic contrast is disentangled into its spectral components (as shown in Supplementary Fig. 11) using Fourier analysis. In **b** and **c** only the long-range contributions to the constant current topography are shown for the areas in a marked with dashed red and blue boxes, respectively. **d** CCT (8 nm x 8nm, $I_T = 0.07\text{nA}$) to the left of a double step on a terrace S3 acquired with a bias voltage of $V_{\text{Bias}} = -0.1\text{V}$, **e** with a bias voltage of $V_{\text{Bias}} = -0.2\text{V}$ and **f** with a bias voltage of $V_{\text{Bias}} = -0.3\text{V}$ representing the integral local density of states in the energy interval $E_F \pm V_{\text{bias}}$. **g** Scanning tunneling spectroscopy at 8K on terraces S2 and S3 separated by a double substrate step as indicated in the constant current topography in the inset ($V_{\text{Bias}} = -0.6\text{V}, I_T = 0.15\text{nA}$). The solid blue line shows the averaging of all spectra recorded on S3 and the solid red line shows the averaging of all spectra recorded on S2.

### Analysis of the local electronic structure

CCTs taken at different bias voltages have additionally been used to gain insight into the local density of states of the combined graphene / interface layer system in a given energy interval $E_F \pm V_{\text{bias}}$. For epitaxial graphene on SiC it is known that for larger voltages $V_{\text{bias}}$ electronic states of the interface layer become visible in CCTs [39]. In Fig. 4d-f we show high-resolution, quasi-simultaneous CCTs recorded at $V_{\text{bias}} = -0.1\text{V}$, $V_{\text{bias}} = -0.2\text{V}$ and $V_{\text{bias}} = -0.3\text{V}$. In all images the graphene honeycomb lattice as well as the $6 \times 6$-quasi corrugation are well resolved. They dominate the topographic contrast at $V_{\text{bias}} = -0.1\text{V}$. In contrast, at higher voltages additional states of the interface are visible as non-periodic defect structures, which is in agreement with published results [39, 40].

Spectroscopic measurements using scanning tunneling spectroscopy (STS) at 8K provide insight into the local electronic structure. dI/dV spectra of graphene on SiC show two prominent minima, firstly the so-called pseudogap at 0 meV and secondly a minimum at the position of the Dirac point [41]. The position of the latter minimum gives a hint to the local charge carrier density [42]. In Fig. 4g the STS data acquired on two terraces connected by a double substrate step are shown. The dI/dV spectra in Fig. 4g are very similar and in agreement with ARPES measurements (Supplementary Fig.

13) and published results [41,43]. Quantitative deviations between STS and ARPES measurements may be due to different measurement conditions such as the temperature. Further factors are the addressability of electronic states in the different techniques and also the presence of the probe itself in STS measurements. In addition to the two prominent minima, we find a pronounced maximum between $-200\text{mV}$ and $-250\text{mV}$, which we assign to the interface states observed in CCT.

On closer inspection of the individual dI/dV spectra it can be seen that the electronic properties on the two terraces are not identical and even on a given terrace we find local deviations (Fig. 4g). In order to quantify these deviations, we describe each individual dI/dV spectrum, in the region of the minimum at negative voltages with a polynomial fit (Supplementary Fig. 14). From the minima of these fits we obtain the position of the Dirac point for each spectrum separately. The variations on a given terrace regarding the position of the Dirac point, are roughly comparable to the differences in dI/dV spectra on the two different terraces (Supplementary Fig. 14). For the terrace to the left we find an average value of $E_D = (-360 \pm 17)\text{meV}$, for the terrace to the right the mean value is $E_D = (-355 \pm 13)\text{meV}$. The error interval here is the standard deviation.

## Discussion

In order to interpret the local transport properties of PASG graphene, we correlate the structural and electronic STM/STS information with the local STP measurement and thereby address a number of questions. Firstly, can we assign the two distinct sheet resistances $\varrho_{\text{Low}}$ and $\varrho_{\text{High}}$ unambiguously to characteristics of the sample? Secondly, what causes the huge spread in the sheet resistance at low temperature found for both $\varrho_{\text{Low}}$ and $\varrho_{\text{High}}$? And finally, can both effects, the differences in $\varrho_{\text{Low}}$ and $\varrho_{\text{High}}$ as well as the spread at low temperature, be traced back to the same origin?

In a first step, we assign the specific step structure revealed in large scale topographies (Fig. 1a) to the stacking sequence of the 6H-SiC(0001) substrate. All SiC crystals consist of fundamental layers of silicon and carbon atoms, arranged in tetrahedral coordination [44, 45, 46]. These layers are referred to as fundamental bilayers. Although the 6H-SiC(0001) exhibits six different (crystal) terminations (labeled as S1, S2, S3 and S1$^*$, S2$^*$, S3$^*$ [47] in Supplementary Fig. 15), only 4 out of the 6 possible 6H-SiC terminations are found [28], because the terraces S1/S1* have a higher decomposition velocity [34, 48] compared to the other terraces and therefore disappear during the growth process. We label the graphene terraces according to the substrate terminations as S2/S2* and S3/S3*. It directly follows, that graphene on terraces S2/S2* exhibits a low sheet resistance and a short-wave spatial modulation of the $6 \times 6$-quasi corrugation. In contrast, a larger sheet resistance $\varrho_{\text{High}}$ and long-wave perturbations of the $6 \times 6$-quasi corrugation are measured on terraces S3/S3*. A systematic difference in $\varrho_{\text{High}}$ for S3 compared to S3$^*$ and in $\varrho_{\text{Low}}$ for S2 compared to S2$^*$ was not observed (Supplementary Fig. 9). Therefore, we refer to S3/S3* as S3 and S2/S2* as S2 in the following (compare Fig. 1c). In summary, we conclude that S2 and S3 are characterized by sheet resistances which differ by their absolute values as well as by their temperature dependence.

We continue our discussion with a more detailed comparison of the local structural and electronic properties of PASG graphene. In general, a variation in the sheet resistance can be caused by a modified charge carrier density, e.g. in the framework of polarization doping [49, 50], as well as a variation in mobility. STS data allow to estimate a difference in the local doping on adjacent terraces. In order to cause the variation in the sheet resistance of 140% for the given terraces, the change in the doping level is expected to become visible in a significant shift of the position of the Dirac point in the STS data of terraces S2 compared to terraces S3. Since the mean Dirac energy on terraces S2 compared to terraces S3 is only shifted by $\approx 5\text{meV}$ (Fig. 4g, Supplementary Fig. 14), we discard a locally varying polarization doping [49, 50] as the main reason for the observed variation of the sheet resistance. Consequently, the local sheet resistance $\varrho$ is predominantly governed by the mobility. The latter is the result of a variety of possible scattering mechanisms like e.g. electron-phonon, electron-electron or electron-defect interaction which all could be modified by the local structural and electronic properties of the sample.

To disentangle possible scattering processes in PASG graphene, we first take the measured step heights into account assuming that they reflect the distance of the graphene layer to the substrate and correlate them with the local transport properties of S2 or S3 (see Fig. 3b and Supplementary Fig. 6e). Data sets like the one presented in Fig. 3b allow for a comparison of two terraces with the same substrate termination, yet different distances of the graphene to the substrate. They directly show that a larger distance of the graphene sheet from the substrate results in a reduced sheet resistance. To further test this hypothesis, we sort the sheet resistances according to the step heights (Supplementary Fig. 5, datasets determined at $8\text{K}$) and, with a single exception, find a match that larger distances result in a reduced sheet resistance. This finding holds for S2 as well as S3 termination.

While the step height variation is not a priori expected, the observed correlation is not surprising. For epitaxial graphene on SiC(0001), the interface layer is partially covalently bonded and thus strongly coupled to the SiC substrate [51], whereas the graphene layer is only weakly coupled [52] by van der Waals interaction. Nevertheless, the electronic properties of epitaxial graphene are known to be strongly influenced by the substrate. Epitaxial graphene shows a strong n-type doping [35, 53, 54] from interface states [49] and a limited charge carrier mobility [55] due to substrate induced scattering [32, 33]. Already a decoupling of the substrate by intercalation leads to an increase in mobility [56], suspended / free-standing graphene shows the highest mobility [13] and a reduced charge carrier density [57]. We suggest that for a larger distance the graphene layer decouples from the substrate resulting in a reduced impact of the defect states of the interface. Thus, these terraces exhibit an increased mobility and a reduced sheet resistance compared to terraces where the graphene layer is closer, i.e. more strongly coupled to the substrate.

Within the proposed model, we now discuss the general trend of the temperature-dependence of the sheet resistance, i.e. an increasing conductivity with decreasing temperature. In the semi-classical Boltzmann transport, a decrease in the sheet resistance (increase in the conductivity) with decreasing temperature is attributed to a dominant phonon scattering contribution [58, 59] and / or electron-electron interaction [60]. The latter contains the competition between the kinetic energy of the individual particles and the scattering among several particles due to Coulomb interaction [60]. It is most dominant at high temperatures and low doping [61]. Since the samples investigated in this study exhibit a strong n-type doping with a high charge carrier concentration of $n \approx 1 \times 10^{13} cm^{-2}$ (Supplementary Fig. 13) we conclude that electron-electron scattering only plays a minor role. From macroscopic magneto transport measurements, we further exclude a dominant impact of phase coherent transport phenomena (see Supplementary Fig. 3b).

Regarding electron-phonon scattering there are two possible origins, firstly intrinsic phonon modes, i.e. acoustic and optical phonon modes of the graphene itself and secondly remote interfacial phonons of the substrate and the interface. As intrinsic optical phonon modes do not couple strongly to the electrons due to their out-of-plane nature [62], electron-phonon scattering is governed by the contribution from remote interfacial phonons [63]. Since the temperature-dependence of the resistance associated with electron-phonon scattering is consistent with our measurements, we attribute a part of the general temperature-dependence to scattering with substrate phonons. Assuming that electron-phonon scattering causes a monotonous decrease of the sheet resistance with decreasing temperature [32, 33], we estimate the phonon contribution $\varrho^{el-phonon}(T)$ as the difference between the mean sheet resistance at 300K and the highest measured values at 8K on terraces S3. This estimation yields a phonon contribution of $< 100\Omega$. Besides the general decrease in the sheet resistance, our data show an increase in the spread of the individual measurements at low temperature (Fig. 2a) which is accompanied by a reduction in the sheet resistance of up to $\approx 250\,\Omega$ when going from 300K to 8K. Within our model, the spread in the data primarily originates from the dependence of the sheet resistance on the distance d to the substrate. From this it directly follows that a local modification of the interaction between the graphene sheet and the substrate results in a locally varying mobility. For electron-phonon scattering, one would expect stronger electron-phonon scattering for smaller d, which does not agree with the observed behavior (Supplementary Fig. 5). This strongly indicates an additional relevant scattering mechanism besides electron-phonon scattering, explicitly depending on d.

Triggered by the observations from CCT, i.e. spatial modulations like the ones observed in Fig. 4a-c, we propose scattering at local defects and potential fluctuations as the additional scattering mechanism: $\varrho(T,d) = \varrho^{el-phonon}(T) + \varrho^{el-defect}(T,d)$. The topographic contrast in highly resolved CCT is dominated by the $6 \times 6$-quasi corrugation. It originates from a lattice mismatch between the SiC and the graphene sheet leading to the development of a hexagonal network of carbon atoms in the interface layer that are not covalently bonded to the substrate [38]. The $6 \times 6$-quasi corrugation is randomly perturbed (Fig. 4) and consequently, each terrace is unique with respect to its defect structure. This deviation from the perfect ordering of the $6 \times 6$-quasi corrugation induces a random potential scattering. A temperature-dependent impact of potential scattering on the resistivity has been studied for charged impurities [64]. At low temperatures, the impact of Coulomb scattering at charged impurities is reduced due to localization of electrons and associated screening. We propose that the basics of this concept can be transferred to our system, replacing the charged defects by the non-periodic potential fluctuations of the $6 \times 6$-quasi corrugation (Fig. 4a-c) as well as interface states (Fig. 4d-f). At low temperatures, electrons are trapped in the random potential, thereby screening the spatial modulations of the potential landscape. This results in a weaker potential landscape for the remaining transport electrons, thus contributing to the general reduction in the sheet resistance.

Furthermore, within the framework of potential scattering it is reasonable that the localization of electrons and associated screening depends on the structural characteristics of the respective terrace, i.e. the exact shape of the random potential landscape. Therefore, the defect structure which is expected to change from terrace to terrace leads to a variation in the potential as well as the screening. This becomes visible as the large spread in the sheet resistance at low temperatures. In addition, we like to mention that at intermediate temperatures (77K) the sheet resistance on S2 already shows a significant spread in the individual measurements (Fig. 2a), whereas a more homogeneous sheet resistance is found for S3.

Having identified the distance between the graphene and the substrate as an important parameter which controls the sheet resistance in general, the question arises whether this parameter also explains the different sheet resistances of the graphene sheet on terraces S2 compared to terraces S3. Fig. 3c shows the difference in the sheet resistance for adjacent terraces S2 and S3 with respect to the experimentally determined step height deviation. It reveals no clear dependence of the variation in the sheet resistance for adjacent terraces S2 and S3 on the height deviation and thus implies that, in addition to the distance dependence of the sheet resistance, intrinsic differences between S2 and S3 exist. Although not providing a comprehensive picture, first indications of these intrinsic differences can be found in the wave length of the structural modulation (Fig. 4) of S2 and S3, in the local defect structure of the interface layer and by comparing the local I(V) spectra acquired on S2 and S3 which show slight differences in the spectra at e.g. -300 mV.

In summary, the spatial homogeneity of PASG graphene allows for a quantitative analysis of electronic transport on a local scale. We have shown a direct correlation of the structural as well as the electronic transport properties with the substrate. In particular, PASG graphene shows a locally inhomogeneous sheet resistance, which is governed by both the substrate termination of the SiC and the distance between the graphene layer and the substrate. A locally varying distance to the substrate is accompanied by a variation of the impact of the interface states such that a larger distance leads to a reduced resistance. By analyzing the temperature-dependence of the sheet resistance we have disentangled different scattering mechanisms and have thereby revealed a large inhomogeneity in the sheet resistance at low temperatures associated with the imperfections of the $6 \times 6$-quasi corrugation and localized defects.

Besides the investigation of fundamental processes in the interaction between substrate and graphene, the reported anisotropy could be exploited in further transport experiments. For example, it is interesting to simply rotate the sample by 90° such that the current is applied parallel to the steps instead of perpendicular. Generally, terraces

S2 with $\varrho_{\text{low}}$ carry more current than terraces S3 with $\varrho_{\text{High}}$ depending on the ratio of the two sheet resistances (see Supplementary Fig. 16) yielding transport channels with a minimum width of about 10 times the Fermi wavelength. Thus, by selecting suitable narrow terraces, quasi 1D electronic transport might be accessible in a 2D sample. In addition, terraces S2 act as nanoscale heat sources and terraces S3 as heat sinks. This enables the investigation of thermal transport, i.e. heat conduction, in low dimensions. Thus, PASG graphene can be a model system to study the interplay between electronic and heat transport with the aim of improving the performance of thermoelectric devices [65]. In this context, the question arises as to the limitations of the reported effect, i.e. a maximum variation in the sheet resistance of 270% at low temperatures. Finally, since terraces S2 and S3 exhibit different electronic and transport properties, it is conceivable to tune the electronic properties of the graphene sheet by specifically changing the ratio of S2 to S3 by an optimized growth process paving the way for improved mobility in epitaxial graphene.

## Methods

### Sample preparation

Graphene samples investigated in this study were grown on the (0001) Si-terminated face of semi-insulating 6H-SiC wafers with small nominal miscut angle of 0.06° towards $[1\bar{1}00]$ direction applying the PASG technique [9, 28]. The idea of this method is to support the growth process with an external carbon source. A polymer is deposited on the substrate using liquid phase deposition before high-temperature sublimation growth is initialized [9, 10, 28]. Samples prepared with this method are almost defect- and bilayer-free and exhibit shallow step heights, as verified in Raman mapping and AFM topographies [9, 28].

**Scanning probe measurements**

The experiments were performed in a custom-built low-temperature STM and in a custom-built room temperature STM under UHV conditions (base pressure <10$^{-10}$ mbar at 300 K, 77 K and 8 K) using electrochemically etched tungsten tips. STS spectra were acquired using standard lock-in technique and a modulation amplitude of $10\mathrm{mV}$. The concept of the STP setup is depicted in Supplementary Fig. 17a. We electrically contact our samples (3 mm x 7 mm) with gold contacts of 50 nm - 100 nm thickness by thermal evaporation in a shadow mask procedure in a two-terminal geometry. In order to eliminate surface contamination, the samples are heated up to 400° C for 30 minutes after reinsertion into the UHV chamber. A voltage $V_\mathrm{cross}$ is applied across the sample via two gold contacts. The voltage $V_\mathrm{STP}(x,y)$, which is a measure of the local electrochemical potential, is adjusted such that the net tunnel current $I_\mathrm{T}$ vanishes and is additionally recorded as a function of position. The resulting potential map (Supplementary Fig. 17b) gives access to the voltage drop along the graphene sheet in the investigated sample area. The simultaneously acquired constant current topography (Supplementary Fig. 17c) allows to directly connect transport and topographic information. The local sheet resistance of each terrace is determined from the potential gradient on the terrace and the current density $j$ as follows [4] $\varrho_\mathrm{sheet} = \frac{dV_\mathrm{STP}}{dx} \cdot \frac{1}{j} = \frac{E_\mathrm{x}}{j}$.

**Finite element simulation with COMSOL**

The local current density $j_\mathrm{local}(x,y)$ was calculated using a finite element simulation based on COMSOL multiphysics® using the AC/DC module. As input parameters we enter the macroscopic ohmic resistance and the global geometry of the sample. Additional topographic information like substrate steps, bilayer regions and corresponding monolayer-bilayer transitions are included according to the structural information from constant current topographies. Step resistivities used in this study are set to $6\,\Omega\mu m, 12\,\Omega\mu m, 18\,\Omega\mu m$ for single, double and triple substrate steps, respectively.

**Data Availability**

The underlying data of this study are available from the authors upon reasonable request.

**References**

[1] Novoselov, K. S. et al., Electric field effect in atomically thin carbon films. *Science* **306** (5696), 666–669 (2004)

[2] Muralt, P., Pohl, D., Scanning tunneling potentiometry*, Appl. Phys. Lett.* **48** (8), 514-516 (1986)

[3] Miccoli, I., Edler, F., Pfnür, H., Tegenkamp, C., The 100th anniversary of the four-point probe technique: the role of probe geometries in isotropic and anisotropic systems, *J. Phys.: Condens. Matter* **27** (22), 223201 (2015)

[4] Willke, P., Schneider, M. A., Wenderoth, M., Electronic Transport Properties of 1D-Defects in Graphene and Other 2D-Systems, *Ann. Phys.* **529** (11), 1700003 (2017)

[5] Ji, S.-H. et al., Atomic scale transport in epitaxial graphene. *Nature Mat.* **11**, 114–119 (2011)

[6] Willke, P., Druga, T., Ulbrich, R. G., Schneider, M. A., Wenderoth, M., Spatial extent of a Landauer residual-resistivity dipole in graphene quantified by scanning tunnelling potentiometry, *Nature Commun.* **6**, 6399 (2015)

[7] Clark, K. W. et al., Spatially Resolved Mapping of Electrical Conductivity across Individual Domain (Grain) Boundaries in Graphene, *ACS Nano* **7**(9), 7956-7966 (2013)

[8] Emtsev, K. V. et al., Towards wafer-size graphene layers by atmospheric pressure graphitization of silicon carbide, *Nature Mat.* **8**, 203-207 (2009)

[9] Kruskopf, M. et al., Comeback of epitaxial graphene for electronics: large-area growth of bilayer-free graphene on SiC, *2D Mater.* **3**, 041002 (2016)

[10] Momeni Pakdehi, D. et al., Homogeneous Large-Area Quasi-Free-Standing Monolayer and Bilayer Graphene on SiC, *ACS Appl. Nano Mater.* **2**, 844-852 (2019)

[11] Dean, C. R. et al., Boron nitride substrates for high-quality graphene electronics, *Nature Nanotech.* **5**, 722–726 (2010).

[12] Riedl, C., Coletti, C., Iwasaki, T., Zakharov, A. A., Starke, U., Quasi-Free-Standing Epitaxial Graphene on SiC Obtained by Hydrogen Intercalation, *Phys. Rev. Lett.* **103**, 246804 (2009)

[13] Bolotin, K. I. et al., Ultrahigh electron mobility in suspended graphene, *Solid State Commun.* **146**, 351-355 (2008)

[14] Ge, J.-L. et al., Weak localization of bismuth cluster-decorated graphene and its spin-orbit interaction, *Front. Phys.* **12** (4), 127210 (2017)

[15] Avsar, A. et al., Spin-orbit proximity effect in graphene, *Nature Commun.* **5**, 4875 (2014)

[16] Song, K. et al., Spin Proximity Effects in Graphene/Topological Insulator Heterostructures, *Nano Lett.* **18**, 2033-2039 (2018)

[17] Haugen, H., Huertas-Hernando, D., Brataas, A., Spin transport in proximity-induced ferromagnetic graphene, *Phys. Rev. B* **77**, 115406 (2008)

[18] Tang, C. et al., Approaching quantum anomalous Hall effect in proximity-coupled YIG/graphene/h-BN sandwich structure, *APL Mater.* **6**, 026401 (2018)

[19] Natterer, F. D., Scanning tunneling spectroscopy of proximity superconductivity in epitaxial multilayer graphene, *Phys. Rev. B* **93**, 045406 (2016)

[20] Wang, Z.-J. et al., Simultaneous n-interaction and n-doping of epitaxial graphene on 6H-SiC(0001) through thermal reactions with ammonia, *Nano Res.* **6**(6), 399-408 (2013)

[21] Clark, K. W. et al., Energy gap induced by Friedel oscillations manifested as transport asymmetry at monolayer-bilayer graphene boundaries, *Phys. Rev. X* **4**(1), 011021 (2014)

[22] Willke, P. et al., Local transport measurements in graphene on $SiO_2$ using Kelvin probe force microscopy, *Carbon* **102**, 470-476 (2016)

[23] Baringhaus, J. et al., Bipolar gating of epitaxial graphene by intercalation of Ge, *Appl. Phys. Lett.* **104**(26), 261602 (2014)

[24] Baringhaus, J. et al., Ballistic bipolar junctions in chemically gated graphene ribbons, *Sci. Rep.* **5** (2015)

[25] Giannazzo, F., Deretzis, I., La Magna, A., Roccaforte, F., Yakimova, R., Electronic transport at monolayer-bilayer junctions in epitaxial graphene on SiC, *Phys. Rev. B* **86**(23), 235422 (2012)

[26] Ella, L. et al., Simultaneous voltage and current density imaging of flowing electrons in two dimensions, *Nature Nanotech.* **14**, 480-487 (2019)

[27] Druga, T., Wenderoth, M., Homoth, J., Schneider, M. A., Ulbrich, R. G., A versatile high resolution scanning tunnelling potentiometry implementation, *Rev. Sci. Instrum.* **81**, 083704 (2010)

[28] Momeni Pakdehi, D. et al., Minimum Resistance Anisotropy of Epitaxial Graphene on SiC, *ACS Appl. Mater. Interfaces* **10**, 6039-6045 (2018)

[29] Landauer, R., Spatial variation of currents and fields due of localized scatterers in metallic conduction, *IBM J. Res. Dev.* **1** (3), 223-231 (1957)

[30] Homoth, J. et al., Electronic transport on the nanoscale: ballistic transmission and Ohm's law, *Nano Lett.* **9** (4), 1588-1592 (2009)

[31] Low, T., Perebeinos, V., Tersoff, J., Avouris, P., Deformation and Scattering in Graphene over Substrate Steps, *Phys. Rev. Lett.* **108**, 096601 (2012)

[32] Jobst, J. et al., Quantum oscillations and quantum Hall effect in epitaxial graphene, *Phys. Rev. B* **81**, 195434 (2010)

[33] Jobst, J. et al., Transport properties of high-quality epitaxial graphene on 6H-SiC(0001), *Solid State Communications* **151**, 1061-1064 (2011)

[34] Yazdi, G. R. et al., Growth of large area monolayer graphene on 3C-SiC and comparison with other SiC polytypes, *Carbon* **57**, 477-484 (2013)

[35] Riedl, C., Coletti, C., Starke, U., Structural and electronic properties of epitaxial graphene on SiC(0001). A review of growth, characterization, transfer doping and hydrogen intercalation, *Journal of Physics D: Applied Physics* **43**(37), 374009 (2010)

[36] Sclauzero, G., Pasquarella, A., Stability and charge transfer at the interface between SiC(0001) and epitaxial graphene, *Microelectronic Engineering* **88**, 1478-1481 (2011)

[37] Kim, S., Ihm, J., Choi, H. J., Son, Y.-W., Origin of Anomalous Electronic Sturctures of Epitaxial Graphene on Silicon Carbide, *Phys. Rev. Lett.* **100**, 176802 (2008)

[38] Berger, C., Conrad, E. H., de Heer, W. A., Epigraphene: Epitaxial Graphene on Silicon Carbide, Physics of Solid Surfaces, Subvolume B, Chiarotti G., Chiaradia, P., Eds, Springer-Verlag Berlin Heidelberg, Vol III/45B (2017)

[39] Rutter, G. M., Imaging the interface of epitaxial graphene with silicon carbide via scanning tunneling microscopy, *Phys. Rev. B* **76**, 235416 (2007)

[40] Cervenka, J., van de Ruit, K., Flipse, C. F. J, Giant inelastic tunneling in epitaxial graphene mediated by localized states, *Phys. Rev. B* **81**, 205403 (2010)

[41] Willke, P. et al., Doping of Graphene by Low-Energy Ion Beam Implantation: Structural, Electronic, and Transport Properties, *Nano Lett.* **15**(8), 5110-5115 (2015)

[42] Joucken, F. et al., Localized state and charge transfer in nitrogen-doped graphene, *Phys. Rev. B: Condens. Matter Mater. Phys.* **85**, 161408 (2012)

[43] Brar, V. W. et al., Scanning tunneling spectroscopy of inhomogeneous electronic structure in monolayer and bilayer graphene on SiC, *Appl. Phys. Lett.* **91**, 122101 (2007)

[44] Schardt J. et al, LEED structure determination of hexagonal α-SiC surfaces, *Surface Science* **337**, 232-242 (1995)

[45] Park, C. H., Byoung-Ho, C., Keun-Ho, L., Chang, K. J., Structural and electronic properties of cubic, 2H, 4H, and 6H SiC, *Phys. Rev. B* **49**(7), 4485-4492 (1994)

[46] Hristu, R., Stanciu, S. G., Tranca, D. E., Polychroniadis, E. K., Stanciu, G. A., Identification of stacking faults in silicon carbide by polarization-resolved second harmonic generation microscopy, *Scientific Reports* **7**, 4870 (2017)

[47] Seyller, T., Passivation of hexagonal SiC surfaces by hydrogen termination, J. Phys.: Condens. Matter 16, S1755 (2004)

[48] Borovikov, V., Zangwill, A., Step bunching of vicinal 6H-SiC{0001} surfaces, *Phys. Rev. B* **79**, 245413 (2009)

[49] Ristein, J., Mammadov, S., Seyller, T., Origin of Doping in Quasi-Free-Standing Graphene on Silicon Carbide, *Phys. Rev. Lett* **108**, 246104 (2012)

[50] Mammadov, S. et al., Polarization doping of graphene on silicon carbide, 2D Materials 1, 035003 (2014)

[51] Emtsev, K. V., Speck, F., Seyller, T., Ley, T., Interaction, growth, and ordering of epitaxial graphene on SiC{0001} surfaces: A comparative photoelectron spectroscopy study, *Phys. Rev. B* **77**, 155303 (2008)

[52] Varchon, F. et al., Electronic Structure of Epitaxial Graphene Layers on SiC: Effects of the Substrate, *Phys. Rev. Lett.* **99**, 126805 (2007)

[53] Berger, C. et al., Ultrathin Epitaxial Graphite: 2D Electron Gas Properties and a Route toward Graphene-based Nanoelectronics, *J. Phys. Chem. B* **108**, 19912-19916 (2004)

[54] Bostwick, A., Ohta, T., Seyller, T., Horn, K., Rotenberg, E., Quasiparticle dynamics in graphene, Nature Physics 3, 36-40 (2007)

[55] Pallecchi. E. at al., High Electron Mobility in Epitaxial Graphene on 4H-SiC(0001) via post-growth annealing under hydrogen, *Scientific Reports* **4**, 4558 (2014)

[56] Speck, F. et al., The quasi-free-standing nature of graphene on H-saturated SiC(0001), *Appl. Phys. Lett.* **99**, 122106 (2011)

[57] Du, X., Skachko, I., Barker, A., Andrei, E. Y., Approaching ballistic transport in suspended graphene, *Nature Nanotech*. **3**, 491-495 (2008)

[58] Hwang, E. H., Das Sarma, S., Screening-induced temperature-dependent transport in two-dimensional graphene, *Phys. Rev. B* **79**, 165404 (2009)

[59] Das Sarma, S., Hwang, E. H., Density-dependent electrical conductivity in suspended graphene: Approaching the Dirac point in transport, *Phys. Rev. B* **87**, 035415 (2013)

[60] Kotov, V. N., Uchoa, B., Pereira, V. M., Electron-Electron Interactions in Graphene: Current Status and Perspectives, *Rev. Mod. Phys.* **84**(3), 1067-1125 (2012)

[61] Mendoza, M., Herrmann, H. J., Succi, S., Hydrodynamic Model for Conductivity in Graphene, *Scientific Reports* **3**, 1052 (2013)

[62] Chen, J.-H., Jang, C., Xiao, S., Ishigami, M., Fuhrer, M. S., Intrinsic and extrinsic performance limits of graphene devices on SiO$_2$, *Nat. Nanotech.* **3**, 206-209 (2008)

[63] Fratini, S., Guinea, F., Substrate-limited electron dynamics in graphene, *Phys. Rev. B* **77**, 195415 (2008)

[64] Stauber, T., Peres, N. M. R., Guinea, F., Electronic transport in graphene: A semiclassical approach including midgap states, *Phys. Rev. B* **76**, 205423 (2007)

[65] Cao, B.-Y., Yao, W.-J., Ye, Z.-Q., Networked nanoconstrictions: An effective route to tuning the thermal transport properties of graphene, *Carbon* **96**, 711-719 (2016)

**Acknowledgements**

A.S. acknowledges financial support by the Deutsche Forschungsgemeinschaft (DFG) through project We 1889/13-1. D.M.P. acknowledges support from the School for Contacts in Nanosystems (NTH nano). P. W. acknowledges support from Institute for Basic Science IBS-R027-D1. This is a pre-print of an article published in Nature Communications. The final authenticated version is available online at: https://doi.org/10.1038/s41467-019-14192-0.

**Author contributions**

A.S., P.W., M.W. planned the experiments. D.M.P. and K.P. prepared the samples. A.S. and G.A.T. performed the room temperature STM measurements, A.S. acquired the low temperature data. A.S. carried out the main part of the data analysis, G.A.T. assisted in the analysis of the room temperature data. A.S. performed the COMSOL simulations. P.S. and F.S. conducted the ARPES measurements. A.S. produced the van der Pauw device and performed the macroscopic transport measurements. A.S. and M.W. wrote the manuscript. All authors discussed the results and commented on the manuscript.

**Competing financial interests.** The authors declare no competing financial interests.

# Supplementary Information: Substrate induced nanoscale resistance variation in epitaxial graphene

**Supplementary Figures**

Supplementary Figure 1

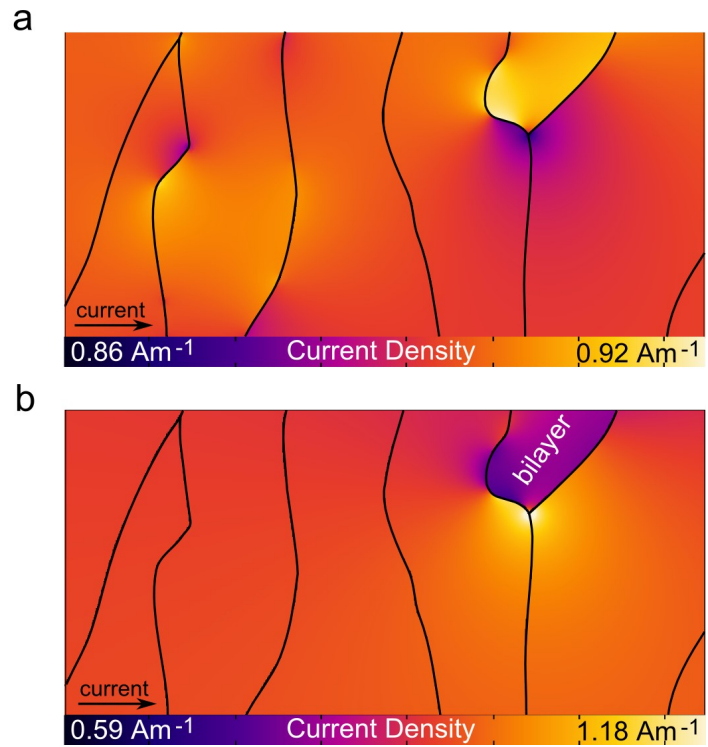

Influence of bilayer regions on the homogeneity of the current density. **a** Using the macroscopic ohmic resistance, the sample geometry and step resistivities of $6\,\Omega\mu m, 12\,\Omega\mu m, 18\,\Omega\mu m$ for single, double and triple steps, respectively, as input parameters, the local current density $j_{\text{local}}(x, y)$ is calculated with finite element simulations using COMSOL for a perfect monolayer grown by PASG. **b** for comparison, a bilayer region with corresponding monolayer-bilayer transition [1] is included in the sample geometry used in a resulting in a highly inhomogeneous local current density. The local variation in the current density increases by almost a factor of ten compared to the pure monolayer case in a as can be seen from the color bars.

Supplementary Figure 2

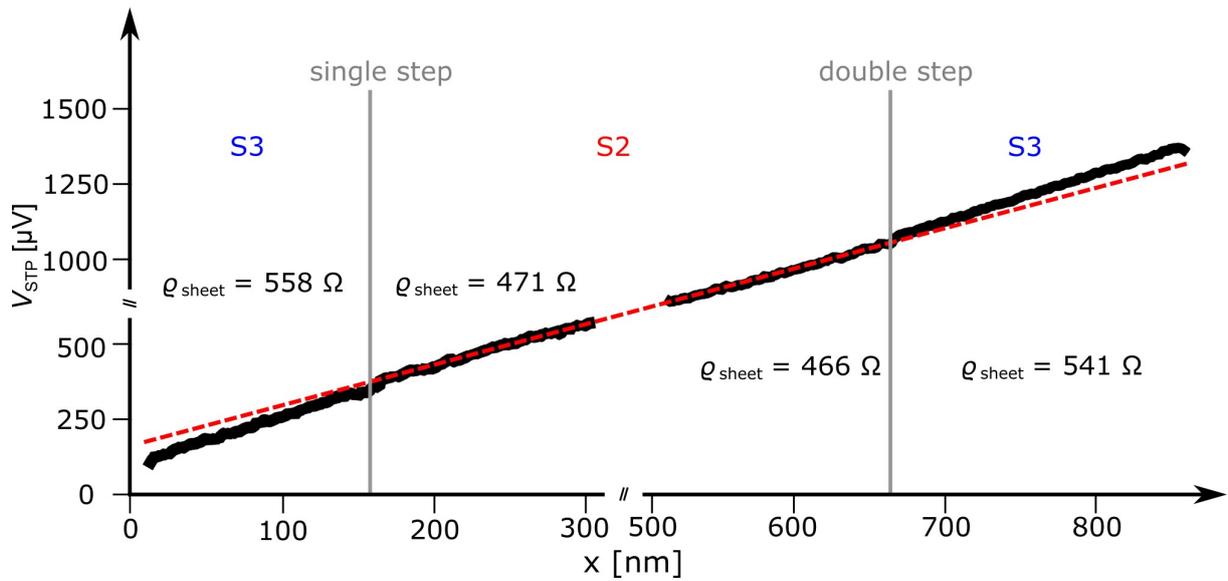

Constant sheet resistance across a given terrace. Voltage drop along the graphene layer for adjacent terraces S3, S2, S3 connected by a single step followed by a double step (left to right). The dashed line represents the slope of the potential in the center region, the variation in the slope in this region is $< 2\%$. Regardless of the measurement position, the sheet resistance can be regarded as constant on a given terrace.

Supplementary Figure 3

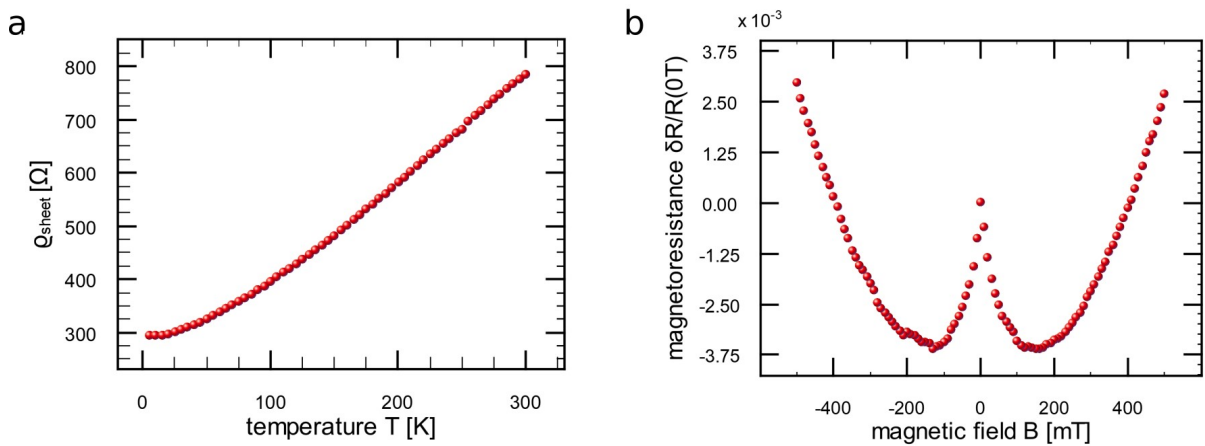

Macroscopic transport measurements in van der Pauw geometry. **a** sheet resistance as a function of temperature in the range of 4K to 300K. **b** magnetoresistance for magnetic fields of $-500$ mT to $+500$ mT acquired at 4K. At small magnetic fields of up to $\pm 100$ mT a negative magnetoresistance is measured, as also observed for conventionally grown epitaxial graphene [2], which then changes to a classical Lorentz magnetoresistance at larger magnetic fields. We attribute the presence of a negative magnetoresistance at small magnetic fields to weak localization. However, the effect of weak localization is significantly less pronounced than in conventionally grown epitaxial graphene [2]. From this we conclude that phase coherent transport phenomena only play a minor role in the samples investigated in this study.

Supplementary Figure 4

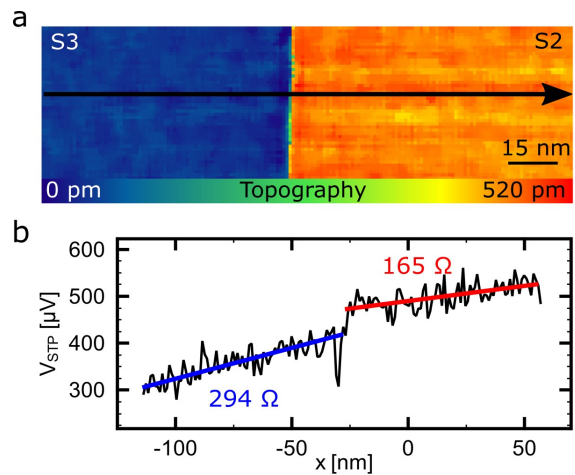

STP data set with a variation in the sheet resistance of 178%. **a** constant current topography (imaging conditions: $V_{\text{Bias}} = 0.03V, I_T = 0.2\ nA, j = 4.07\ \text{Am}^{-1}$) of monolayer graphene crossing a double substrate step. **b** averaged potential along the black arrow, solid red and blue lines indicate the slope of the potential from which the sheet resistance is calculated.

Supplementary Figure 5

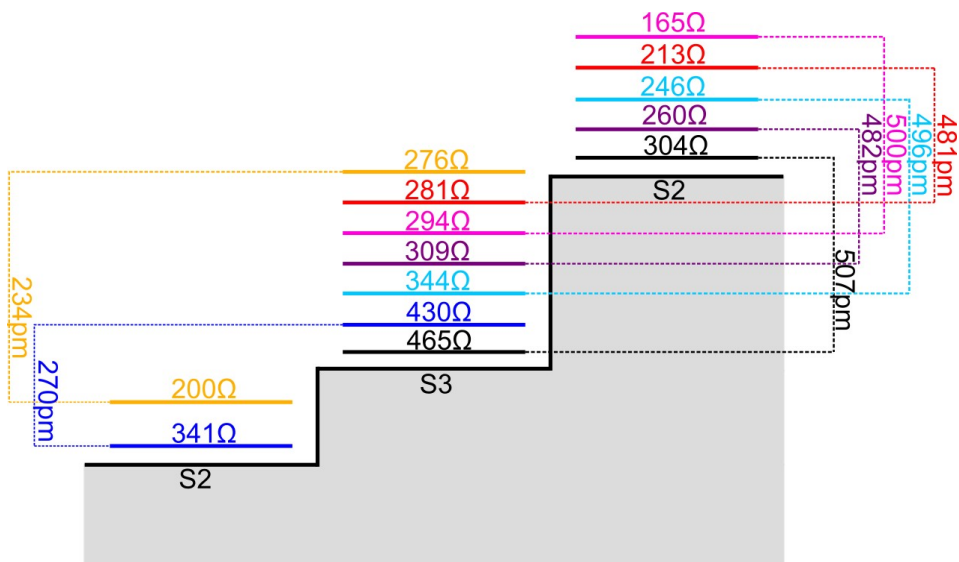

Sorting of the data acquired at $8K$ under the assumption that a larger distance to the substrate leads to a reduction of the resistance. For each terrace, the measured sheet resistances are arranged such that for larger values the distance to the substrate decreases (not to scale). The dotted lines connect adjacent terraces and indicate the measured step height. By comparing different data sets, predictions for the step height can be made. The pink data set exhibits a step height of $500$pm. Compared to the pink data set, the red data set shows a lower sheet resistance on terrace S3 and a higher sheet resistance on terrace S2. Thus, according to the proposed model, a step height $< 500$pm is expected for the red data set, which agrees with the measured step height of $481$pm. The only exception is the yellow data set.

Supplementary Figure 6

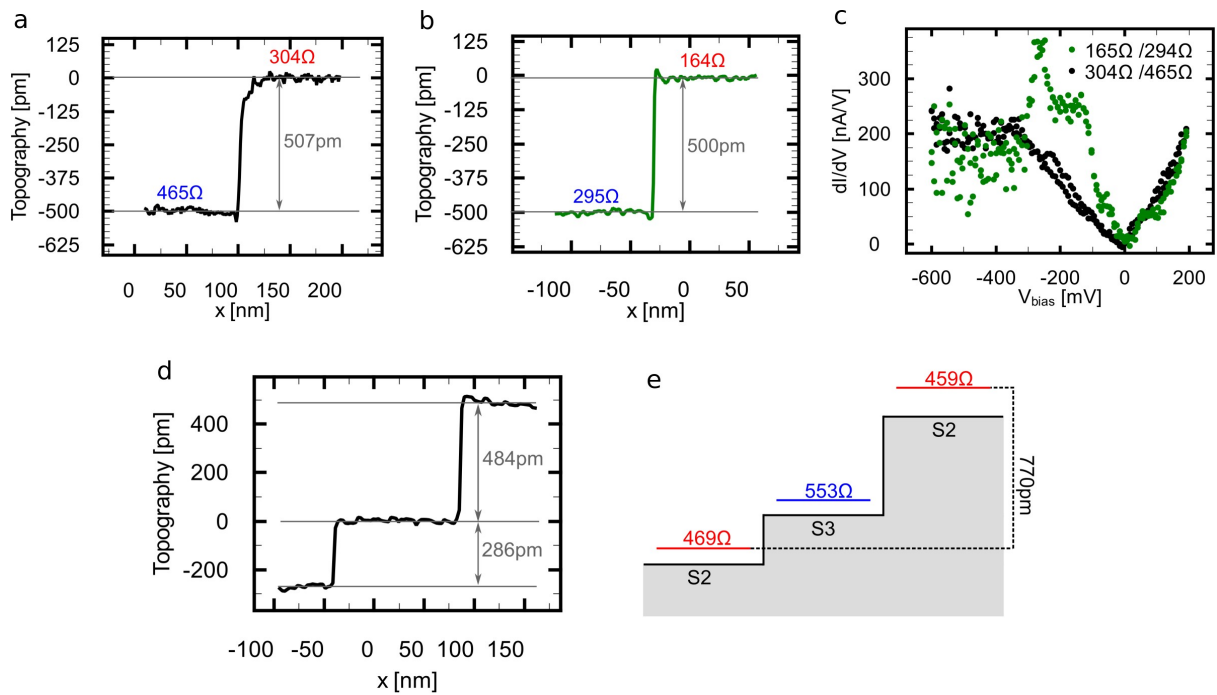

Sheet resistance and step height. **a** sheet resistance and step height for the largest sheet resistance measured at 8K **b** sheet resistance and step height for the smallest sheet resistance measured at 8K. **c** Scanning tunneling spectroscopy corresponding to the data sets shown in a and b. **d** Line profile through a constant current topography ($V_{\text{Bias}} = -0.03\text{V}, I_{\text{T}} = 0.2\text{nA}$) showing adjacent terraces S2, S3, S2, connected by a single substrate step followed by a double substrate step recorded at 300K. The line profile reveals a deviation from the step heights of the SiC substrate steps. **e** schematic representation of the correlation between step height and sheet resistance illustrating a locally varying distance between the graphene layer and the substrate.

Supplementary Figure 7

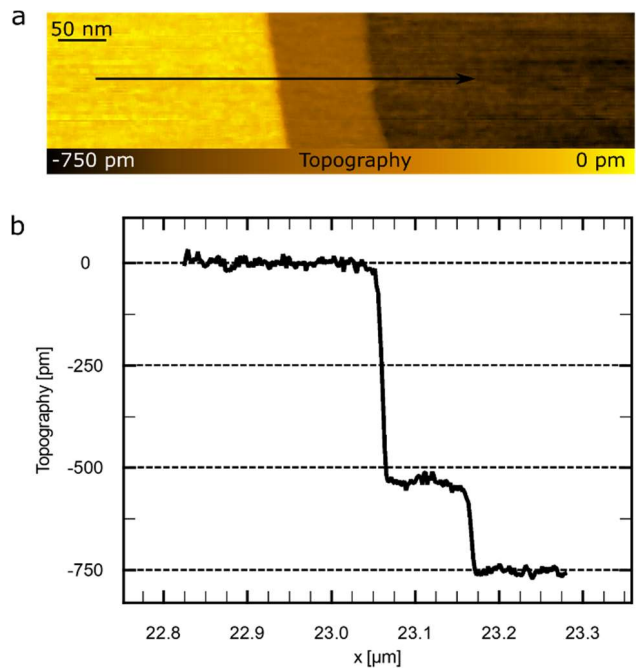

Topographic analysis using atomic force microscopy. **a** AFM topography and **b** line profile along the black line in a reveal a step height $< 0.25$ nm for the single substrate step and a step height $> 0.5$ nm for the double substrate step.

Supplementary Figure 8

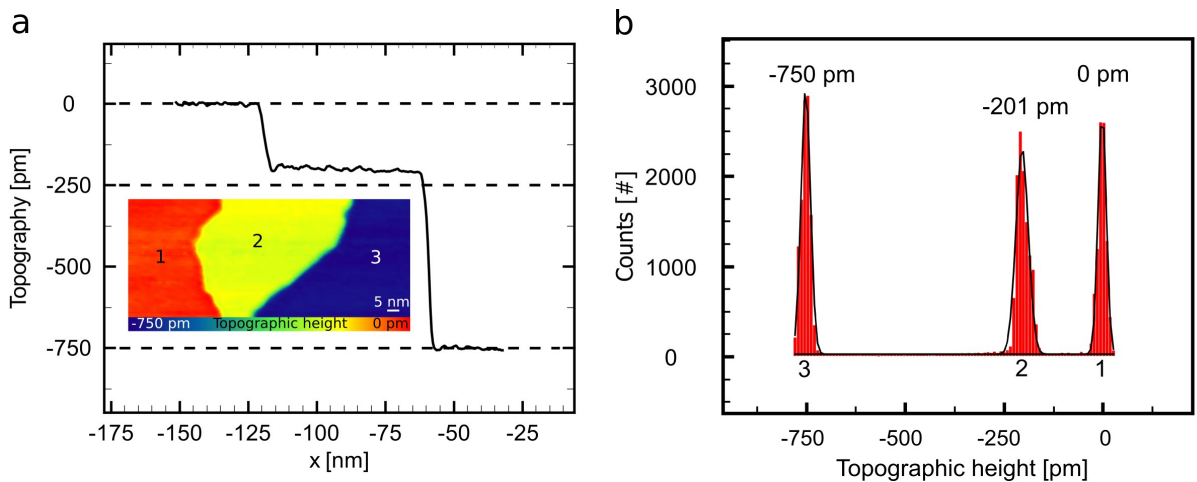

Step height analysis using a histogram method. **a** Line profile through a constant current topography (inset: 600 nm x 100 nm, $V_{\text{Bias}} = -0.03\text{V}, I_{\text{T}} = 0.2\text{nA}$) showing adjacent terraces S3, S2, S3, connected by a single substrate step followed by a double substrate step. **b** height analysis based on evaluating the height information of each pixel. Gaussian curves are fitted to the peaks, the center position of the individual fits are denoted.

Supplementary Figure 9

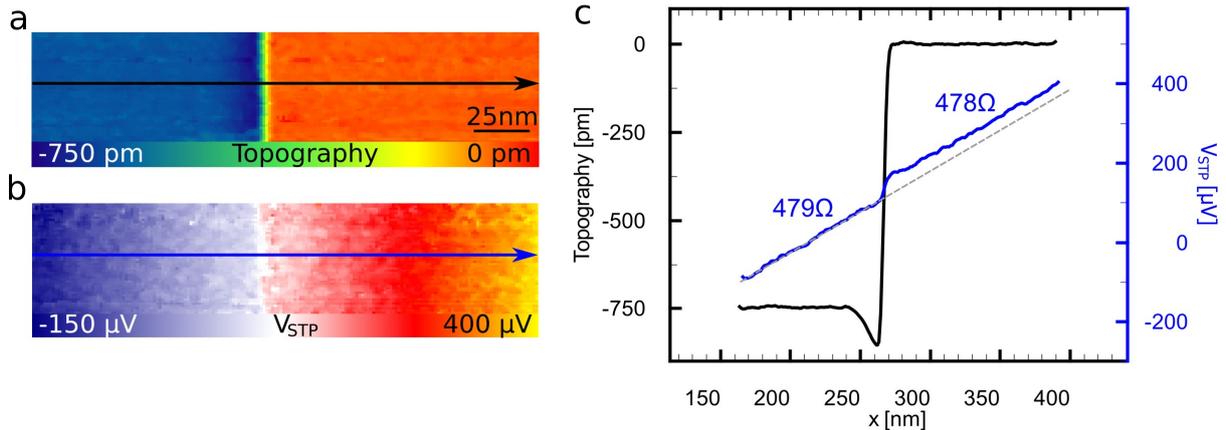

Height calibration using a triple substrate step with almost identical sheet resistance to the left and to the right. **a** constant current topography (imaging conditions: $V_{\text{Bias}} = 0.03V, I_{\text{T}} = 0.2\,nA, j = 4.08\,\text{Am}^{-1}$) of monolayer graphene crossing a triple substrate step, **b** simultaneously recorded potential map. **c** topographic height averaged along the black arrow in a and averaged potential along the blue arrow in b. The fact that triple steps show a step height of 750pm confirms the correct height calibration of the piezo. The calibration was checked for all examined temperatures at several triple substrate steps.

Supplementary Figure 10

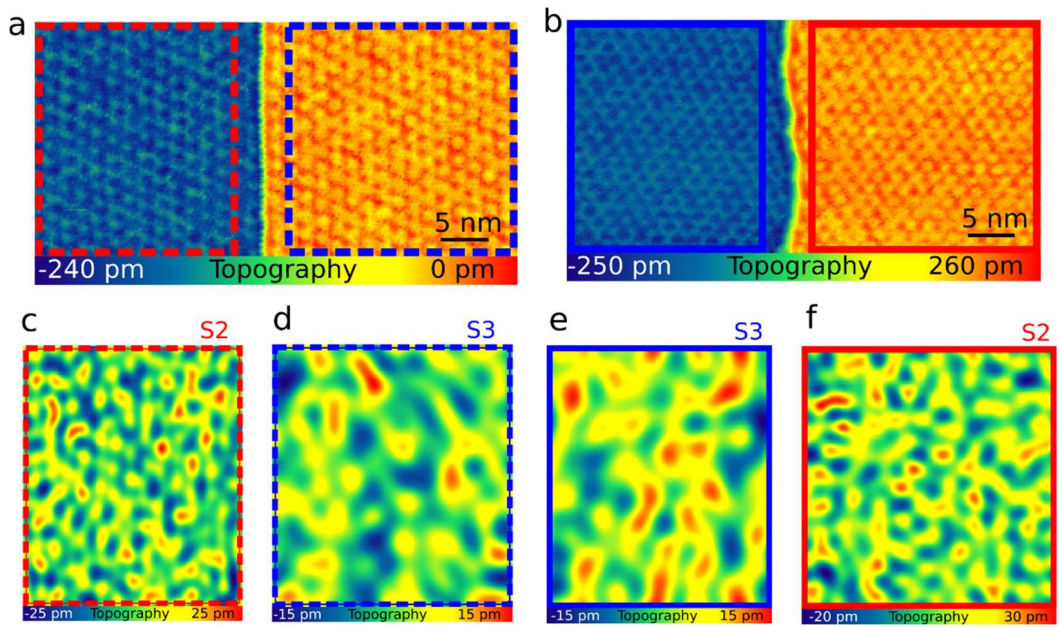

Analysis of the topographic contrast on terraces S2 and S3. **a** 50 nm x 25 nm constant current topography of terraces connected by a single substrate step and **b** connected by a double step ($V_{\text{Bias}} = -0.3V, I_{\text{T}} = 0.15\text{nA}$). On all four terraces the $6 \times 6$ modulation is well resolved. The topographic contrast is disentangled into its spectral components (as shown in Supplementary Fig. 11) using Fourier analysis. In **c** and **d** only the long-range contributions to the constant current topography are shown for the areas in b marked with dashed red and blue boxes, respectively. **e** and **f** depict the corresponding long-range contributions for the areas in c marked with solid red and blue boxes.

Supplementary Figure 11

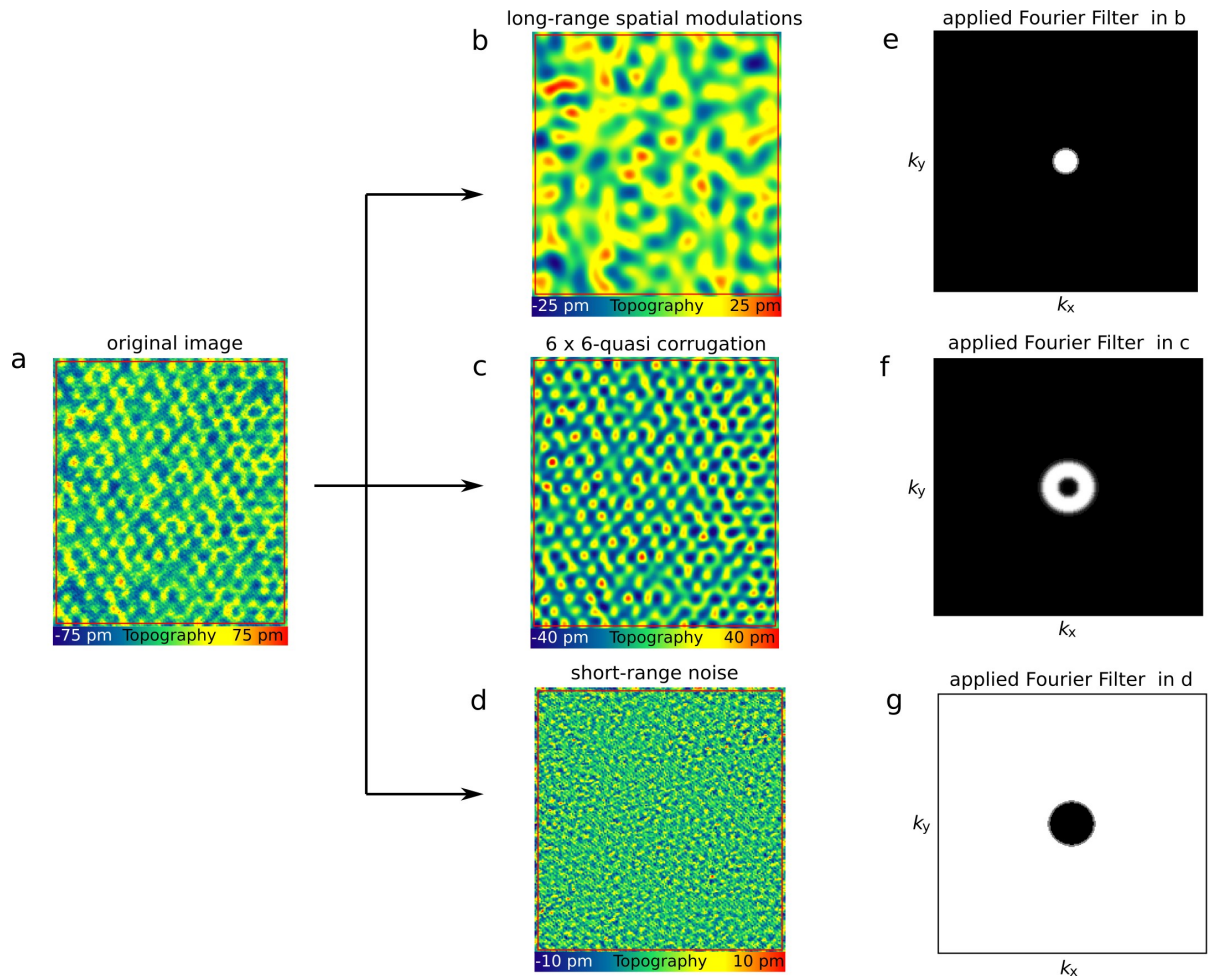

**Spectral disentanglement of constant current topographies. a** original image is disentangled into its spectral components using different Fourier filters: **b** long-range spatial modulation, **c** the 6 × 6-quasi corrugation, **d** and short-range noise. **e, f** and **g** applied Fourier filters in b, c and d, respectively. Dark regions indicate spectral components that are filtered out.

Supplementary Figure 12

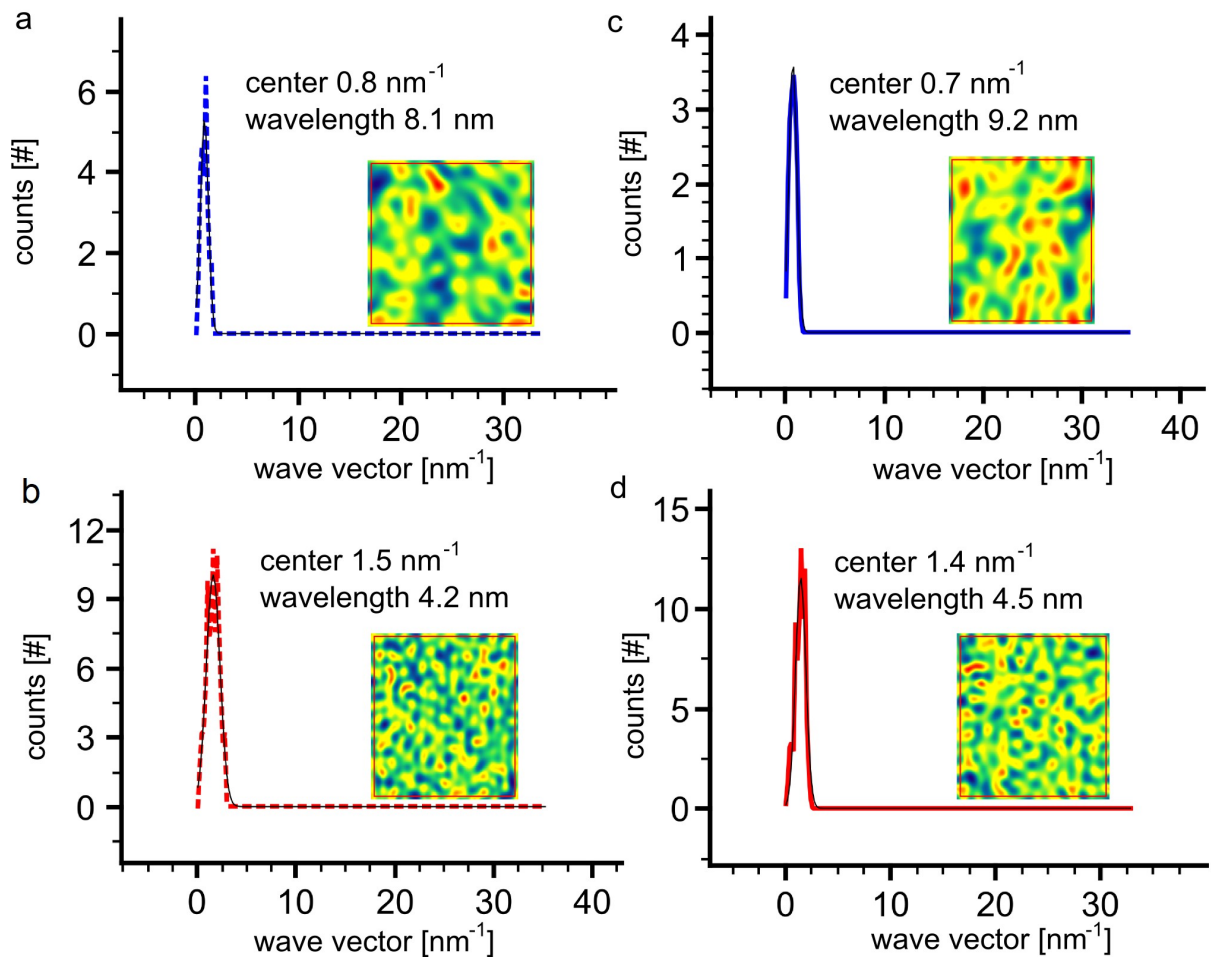

Spectral analysis of the long-range spatial modulations. **a** spectral analysis of the upper terrace in Fig. 4a, **b** spectral analysis of the lower terrace in Fig. 4a. **c,d**, spectral analysis of the topographic contrast in Supplementary Fig. 10b. The original CCTs are Fourier filtered as shown in Supplementary Fig. 11. The resulting long-range contributions are converted into powerspectra for each terrace separately. The wavelength of the spatial modulation is calculated from the center position of a Gauss-Fit adjusted to each powerspectrum.

## Supplementary Figure 13

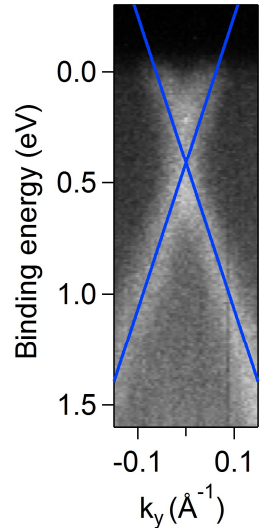

ARPES measurements of the $\pi$-bands near $E_F$ at the K-point of the graphene Brillouin zone. The photon energy was $\hbar\omega = 40.81$ eV. The blue lines correspond to fitted tight-binding bands and the resulting Dirac energy is $E_D - E_F = 410$ meV.

## Supplementary Figure 14

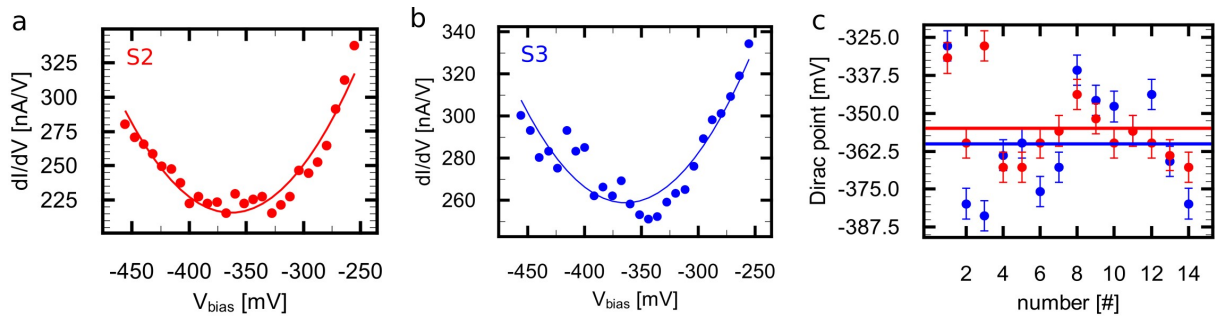

Evaluation of the position of the Dirac point. **a** close-up of a single spectrum recorded on a terrace S2 in the voltage range of $-250$mV to $-450$mV. The solid line shows a polynomial fit. The position of the Dirac point is given by the position of the minimum of the polynomial fit. **b** close-up of a single spectrum acquired on a terrace S3 and corresponding fit. **c** determined Dirac points for all dI/dV spectra shown in Fig. 4g. On terraces S2 we find an average value of $E_D^{S2} = (-355 \pm 13)$meV, on terraces S3 the mean value is $E_D^{S3} = (-360 \pm 17)$meV as indicated by the solid lines. The denoted error interval is the standard deviation.

Supplementary Figure 15

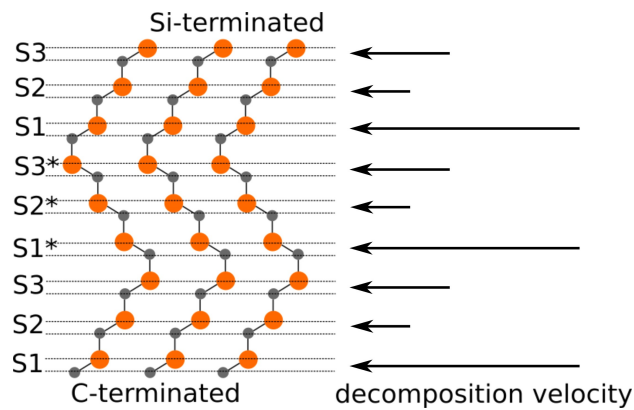

Crystal structure of 6H-SiC(0001). Schematic side view of the crystal structure of 6H-SiC and the decomposition velocity according to [3]. For the decomposition velocities, there are different conclusions in literature as to whether terraces S2/S2* or terraces S3/S3* show a higher decomposition velocity, compare [3] and [4]. However, there is agreement that S1/S1* are the terraces with the highest decomposition velocity and thus disappear first during the growth process.

Supplementary Figure 16

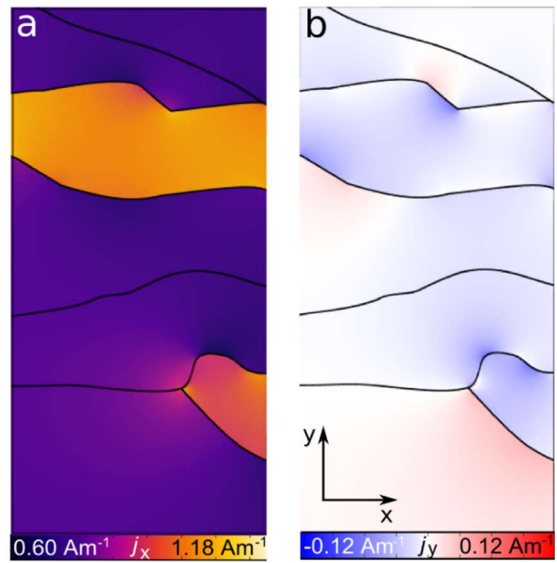

Intrinsic quasi 1D current channels. **a** x-component of the current density from finite element simulation with current flow parallel to the substrate steps. In addition to the sample geometry and the step resistances, each terrace has been assigned a sheet resistance $\varrho_{\text{sheet}}$ according to the underlying SiC crystal surfaces. **b** y component of the current density.

Supplementary Figure 17

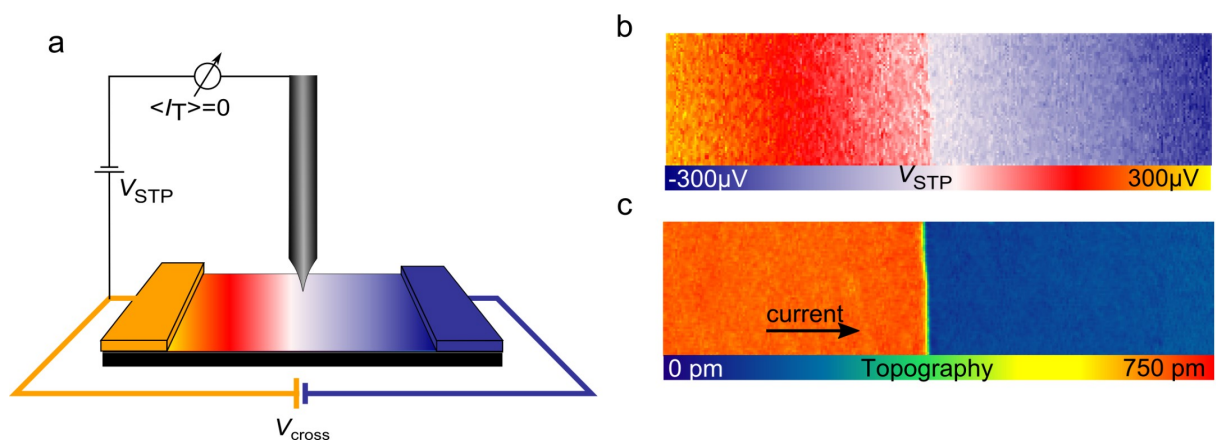

Working principle of our STP setup. **a** schematic drawing of the STP setup: a graphene sample is contacted in two-terminal geometry and a voltage $V_{\text{cross}}$ is applied across the sample. The voltage $V_{\text{STP}}(x,y)$ is adjusted such that the net tunnel current $I_T$ vanishes. It is recorded at every position of the topography and represents the electrochemical potential of the sample at the position of the tip. **b** resulting potential map and **c** simultaneously recorded $(200 \times 50)\,\text{nm}^2$ topography (imaging conditions: $V_{\text{Bias}} = 0.03\,\text{V}, I_T = 0.15\,\text{nA}, j = 3.56\,\text{Am}^{-1}$) of monolayer graphene crossing a triple substrate step.

**Supplementary Tables**

| measurement | 1 | 2 | 3 | 4 | 5 | 6 | 7 |
|---|---|---|---|---|---|---|---|
| current density [Am$^{-1}$] | 0,880 | 0,885 | 0,885 | 0,880 | 0,882 | 0,882 | 0,891 |

**Supplementary Table 1 | Evaluation of the current density.** Current densities for all marked areas in Fig. 1a (from left to right) determined from finite element simulations. The macroscopic average current density is $j = 0.89\,\text{Am}^{-1}$ per applied volt cross voltage $V_{\text{cross}}$.

**Supplementary References**

[1] Willke, P., Druga, T., Ulbrich, R. G., Schneider, M. A., Wenderoth, M., Spatial extend of Landauer residual-resistivity dipole in graphene quantified by scanning tunnelling potentiometry, *Nature Commun.* **6**, 6399 (2015)

[2] Willke, P. et al., Doping of Graphene by Low-Energy Ion Beam Implantation: Structural, Electronic, and Transport Properties, *Nano Lett.* **15**(8), 5110-5115 (2015)

[3] Yazdi, G. R. et al., Growth of large area monolayer graphene on 3C-SiC and comparison with other SiC polytypes, *Carbon* **57**, 477-484 (2013)

[4] Borovikov, V., Zangwill, A., Step bunching of vicinal 6H-SiC{0001} surfaces, *Phys. Rev. B* **79**, 245413 (2009)